# Simultaneously improving multiple imaging parameters with scattering media

**Fu Zhao,**[1, 2,] **Shuman Du,**[1,2,] **Dong Liang**[1,3], **Jun Liu,**[1,2,*]

[1] *State Key Laboratory of High Field Laser Physics and CAS Center for Excellence in Ultra-intense Laser Science, Shanghai Institute of Optics and Fine Mechanics, Chinese Academy of Sciences, Shanghai 201800, China*
[2]*University Center of Materials Science and Optoelectronics Engineering, University of Chinese Academy of Sciences, Beijing 100049, China*
[3]*School of Physics Science and Engineering, Tongji University, Shanghai 200092, China*

*\*jliu@siom.ac.cn*

**Abstract:** Traditional optical imaging systems can provide high-quality imaging with complicated and expensive optical design by eliminating aberrations. With the help of optical memory effect (ME), not independently improving single imaging parameter, but simultaneously improving several imaging parameters by adding scattering media to the imaging systems was demonstrated firstly. As an example, in a simple single-lens imaging system, except the depth-of-field (DOF) was improved greatly, spherical aberration, coma aberration and chromatic aberration were eliminated simultaneously by placing a scattering medium between the lens and the camera. The results indicate the potential applications of scattering media in many fields such as optical imaging, optical measurements and biomedical applications.

## 1. Introduction

In an optical imaging system, improving image quality by eliminating optical aberrations has always been the ultimate goal of researchers. With the development of the ray tracing technology and corresponding algorithms based on geometric optics [1-5], the optimization task of the optical imaging system can now be effectively performed, where typical optical aberrations are spherical aberration, coma aberration, astigmatism, and chromatic aberration [6]. Eliminating one kind of aberration can only partially improve the imaging quality, sometimes even at the cost of other parameters. Realizing a global optimization of optical imaging systems requires all aberrations to be corrected, which means an extremely complex optical design, and the optical engineering task also suffer from many challenges [7].

Due to the multiple scattering of light in the scattering media, imaging through scattering media will become speckle pattern which is hard to be recognized. Recently, it was found that two-dimensional image information of the object can be directly recovered from the speckle pattern based on the optical memory effect (ME)[8, 9]. Many efforts have been demonstrated to achieve imaging through scattering media thereafter. It was discovered that imaging together with scattering media also show several interesting properties: increasing numerical aperture [10], extending the depth-of-field (DOF) or field-of-view (FOV) of image system [11-14], and enhancing spatial resolution [15-17]. According to the property of increasing numerical aperture, Choi et al. developed a method of imaging by placing the scattering medium between the object and the lens, which overcome the diffraction limit of a conventional imaging system [10]. By using a high refractive index scattering medium as an imaging lens for fluorescence imaging, Yilmaz et al. not only enhanced the resolution but also achieved a wide FOV [12]. A new coherent imaging technique, termed ptychographic structured modulation (PSM), for quantitative microscopy was reported by Song et al. they placed a thin scattering medium in between the sample and the objective lens to modulate the complex light waves from the object, and achieved a 4.5-fold resolution gain over the diffraction limit [17]. Imaging through scattering media requires spatially incoherent illumination, generally narrow-band laser light sources are used, but experiments have proven that broadband illumination can be used without appreciably affecting the performance of imaging [18, 19].

In the paper, we demonstrate simultaneous several imaging parameters improvement in a single-lens imaging system by adding a scattering medium directly after the lens. In the experiment, the speckle patterns of object behind a scattering medium were captured by camera and reconstructed by phase retrieval algorithms. Owing to the random phase caused by scattering media, important imaging parameters such as DOF was improved greatly, moreover, optical aberrations such as spherical aberration, coma aberration and chromatic aberration can be eliminated simultaneously in comparison to the imaging system without scattering medium. As a result, multiple imaging parameters can be improved at the same time according to the experimental results. It means that the scattering medium is a simple optical element which owe the ability to improve the global imaging capabilities of an image system.

## 2. Principle and Method

The light will be scattered when it propagates through scattering media, which diffuses the optical beam into a speckle pattern and scrambles the spatial information [20, 21]. According to the ME, lots of the wavefront information included in a little angle when the light propagates through a scattering medium [22]. Within the angular optical memory effect range of the scattering medium, the speckle pattern $I$ is the convolution between the object $O$ and the system point spread function (PSF) $S$

$$I = \int_{-\infty}^{\infty} O(r) S(r - \Delta r) d^2 r = O * S \tag{1}$$

where $*$ denotes the convolution operation.

The object can be separated from the speckle pattern by autocorrelation. Here, we calculate the autocorrelation of the detected speckle pattern:

$$[I \otimes I] = [O * S] \otimes [O * S] = [O \otimes O] * [S \otimes S] \tag{2}$$

where $\otimes$ denotes the autocorrelation operation.

Assuming that the speckle pattern is randomly distributed in space, the autocorrelation of itself will be an impulse function, i.e., $[S \otimes S] = \delta(r)$. As a result, we can obtain that

$$[I \otimes I] \approx [O \otimes O] \tag{3}$$

The autocorrelation of the object can be numerically inverted using phase iterative algorithm to yield a high-resolution image. According to the convolution theorem,

$$F\{I \otimes I\} = F\{O \otimes O\} = F\{O\} F\{O\}^* = |F\{O\}|^2 \tag{4}$$

As a result, the retrieval of the correct phase in the Fourier domain will lead to the reconstruction of the object which was hidden behind the scattering medium. In this study, we adopted two phase retrieval algorithms, called the error reduction algorithm and the hybrid input-output algorithm, Here, we adopted ping-pang (PP) algorithm [23], which is a mixture of the error reduction algorithm and the hybrid input-output algorithm, to reconstruct the object with the high-speed and high-quality.

## 3. Experimental setup

In order to verify that scattering medium can improve the depth-of-field (DOF)，especially is able to eliminate spherical aberration, coma aberration, and chromatic aberration at the same time, experiments with a scattering diffuser in a single lens imaging system have been carried out for simplicity and generality. The schematic of experimental setup is shown in Figure. 1. Spatial incoherent illumination is composed of two collimating lenses ($L_1$ and $L_2$), a rotating diffuser (RD, DG20-600, Thorlabs) with a rotation speed of 20Hz and a fiber coupled multi-wavelength available laser (RGB-637/532/450/405nm-160mW-177021) as excitation light source that can emit light at 637/532/450/405nm 4 different wavelengths. The object in the experiment is a digital number '2' on the resolution target (Negative 1951 USAF Test Target, ThorLabs), and it is fixed on a two-dimensional translation stage that can move in the directions of the optical axis and perpendicular to the optical axis. The imaging lens (L3, the focal length $f$=100 mm) is also placed on a one-dimensional translation stage. The scattering medium is a special rough glass plate (DG20-220, ThorLabs), which is placed adjacent to the other side of the imaging lens. The transmission light after the scattering medium is captured by a sCMOS camera (Dhyana 401DS, Tucsen with 2048×2048 pixels), which is placed directly after the scattering medium. For a conventional single lens imaging, the object distance $d_o$ and image distance $d_i$ are set to be 300mm and 150mm respectively, which follow the imaging law of a thin lens, which is $1/f = 1/d_o + 1/d_i$.

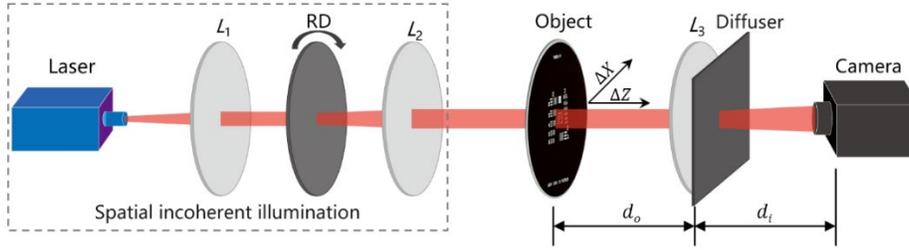

Figure. 1. Schematic of experimental setup. L1-L3: optical lens, RD: a rotating diffuser, Object: Negative 1951 USAF Test Target, Diffuser: the scattering medium.

## 4. Experimental results and discussion

Firstly, we demonstrated that the DOF can be greatly extended by using scattering media. In the experiment, the image distance was kept as a constant ($d_i$ = 150 mm), while the object distance was changed. In the absence of the scattering medium, a clear image (Fig. 2. c1) could be obtained when the object was placed at the focus plane (ΔZ = 0 mm). However, when the object moved along the optical axis, either close to or away from the lens for 5 mm (ΔZ = -5 and 5 mm), the images became blurred (Fig. 2. b1 and d1). As the distance increased to 10 mm (ΔZ = -10 and 10 mm), the images would be further blurred and even fail to identify (Fig 2. a1 and e1). If a scattering medium is located between the lens and the camera, all the images at different DOF were shown with speckle patterns (Fig. 2. a2-e2). After a series calculation based on the phase-retrieval algorithm, clear images (Fig. 2. a3-e3) could be obtained from all the speckle patterns. Simply adding a scattering medium, clear object images with different DOF can be obtained from the speckle patterns, which means the scattering medium can be used to extend the DOF of an imaging system.

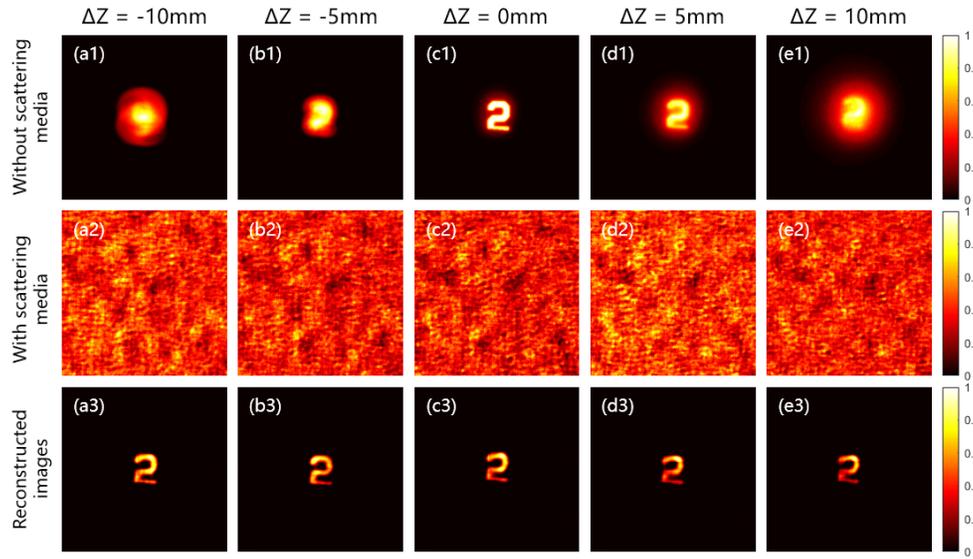

Figure. 2. Results at different DOF with or without scattering medium. (a1-e1) Direct images without scattering medium. (a2-e2) Speckle patterns with scattering medium. (a3-e3) Images reconstructed from speckle patterns by using phase-retrieval.

In a traditional optical imaging system, when the object is far away from the optical axis, there will be coma aberration. In the next experiment, the object was fixed on a laterally moving translation stage that can move it perpendicular to the optical axis, while the image distance $d_i$ and the object distance $d_o$ remain unchanged. When the object was moved by 2 mm (ΔXo = -2 and 2 mm) laterally, one side of the images (Fig. 3. b1 and d1) captured by the camera began to blur due to coma aberration. With the increasing of the shifted distance to 5 mm (ΔXo = -5 and 5mm), which is far from the paraxial condition, the images (Fig. 3. a1 and e1) were further blurred and the brightness of the images were reduced. Again, when the scattering medium was inserted into the imaging system, clear images could be recovered from all the captured speckle patterns. The lateral moving distance was limited by the radius of the collimator lens ($R_{L_2}$ = 12.7mm), which cannot be large in our device, and there will be no light on the target when the movement distance exceeded the radius of the collimating lens.

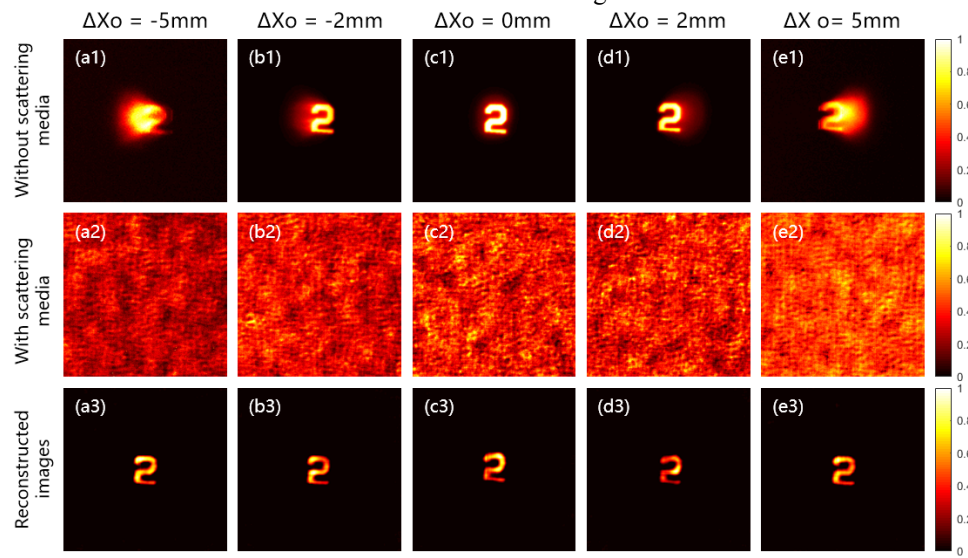

Figure. 3. Results of moving object perpendicular to the optical axis with or without scattering medium. (a1-e1) Direct images without scattering medium. (a2-e2) Speckle patterns with scattering medium. (a3-e3) Images reconstructed from speckle patterns by using phase-retrieval.

In order to introduce spherical aberration in the imaging system, we placed the imaging lens on a translation stage, which can make it move perpendicular to the optical axis, while other optical components and parameters remain unchanged. When the lens moved by 4 mm at both sides (ΔXl = -4 and 4mm) where the image almost moved out of the camera's imaging range, a noticeable blur appears (Fig. 4. a1 and e1). Again, with the scattering medium, the original images can be recovered from every obtained speckle patterns.

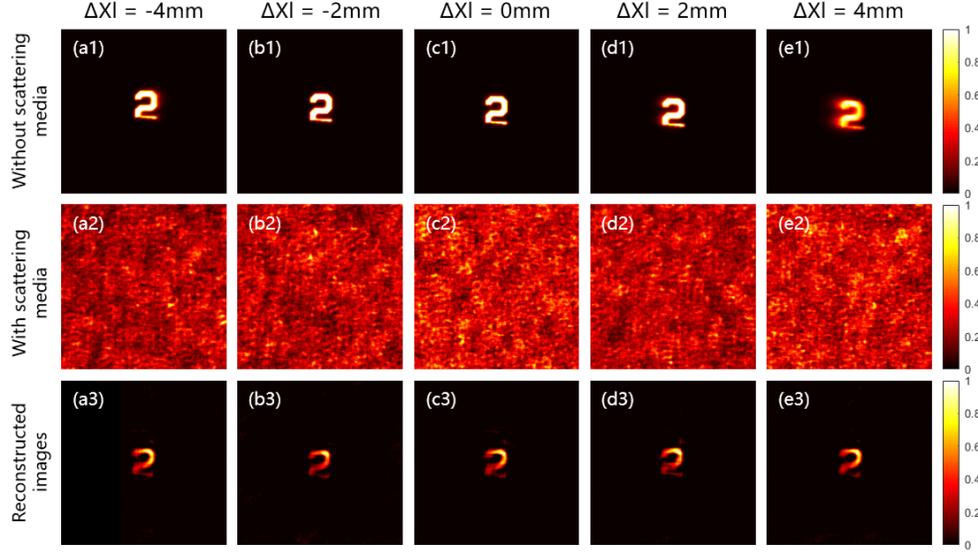

Figure. 4. Results of moving imaging lens perpendicular to the optical axis with or without scattering medium. (a1-e1) Direct images without scattering medium. (a2-e2) Speckle patterns with scattering medium. (a3-e3) Images reconstructed from speckle patterns by using phase-retrieval.

The chromatic aberration is another important factor that limits the imaging capabilities of imaging systems. Different wavelengths of light have different focal positions after passing through a lens, which will cause the final image to be blurred. A biconvex lens (N-BK7, Thorlabs. Design wavelength is 633nm) was used for imaging in the experiment. The object distance $d_o$ and image distance $d_i$ remained at 300mm and 150mm, respectively. When the wavelength of the light source was 637nm, clear images (Fig. 5. a1 and a3) can be captured by the camera. As the wavelength of the light source switched to 532nm and 405 nm, the focal length of the lens decreased and the captured pictures become more and more unclear due to out of focus or chromatic aberration. Multispectral illumination also made the images tend to be blurred and not clear. Once again, a scattering medium was added between the lens and the camera. In order to eliminate the influence of light intensity of different wavelengths, the light intensities at all wavelengths were adjusted so that the average intensities of all the images through the scattering medium were the same. In all cases, the objects could be clearly recovered from the different speckle patterns. Therefore, it means imaging through a scattering medium can be used to eliminate chromatic aberration in the imaging system.

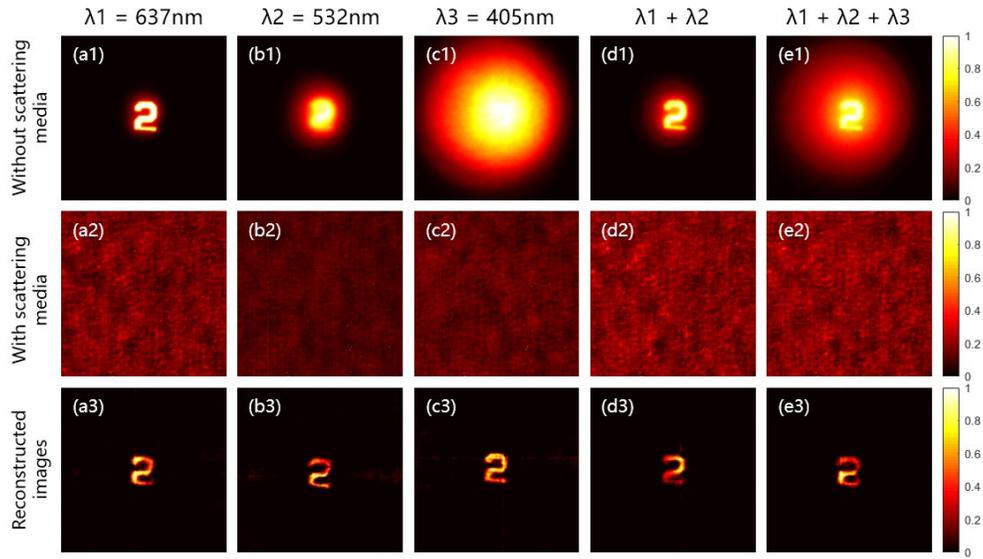

Figure. 5. Results of different wavelengths illumination with or without scattering medium. (a1-e1) Direct images without scattering medium. (a2-e2) Speckle patterns with scattering medium. (a3-e3) Images reconstructed from speckle patterns by using phase-retrieval.

All the above experiments were carried out independently to verify that scattering media can remove the influence of several single aberration in the imaging systems. Can all these aberrations be eliminated simultaneously by using simple scattering media, which otherwise need complicated and expensive optical design? Then, imaging experiments with four aberrations of defocus, coma, spherical and chromatic aberrations included simultaneously were done. The relevant results are shown in Fig. 6. The first column in Fig. 6 shows the results of the imaging system with various aberrations at the same time. A fuzzy number "2" can be seen in Fig. 6. a1 and d1, because for the lens, the focal length of different wavelengths is different, and the object was just on the imaging plane of the appropriate wavelength. It was exciting to see that clear objects can be recovered from both the speckle patterns with the same phase-retrieval process, as shown in Fig. 6. a3 to d3. It means that the scattering medium can be used to simultaneously eliminate the influence of multiple aberrations of imaging systems. This simple method will extend multiple imaging parameter simultaneously with a scattering medium, which would be useful in many applications.

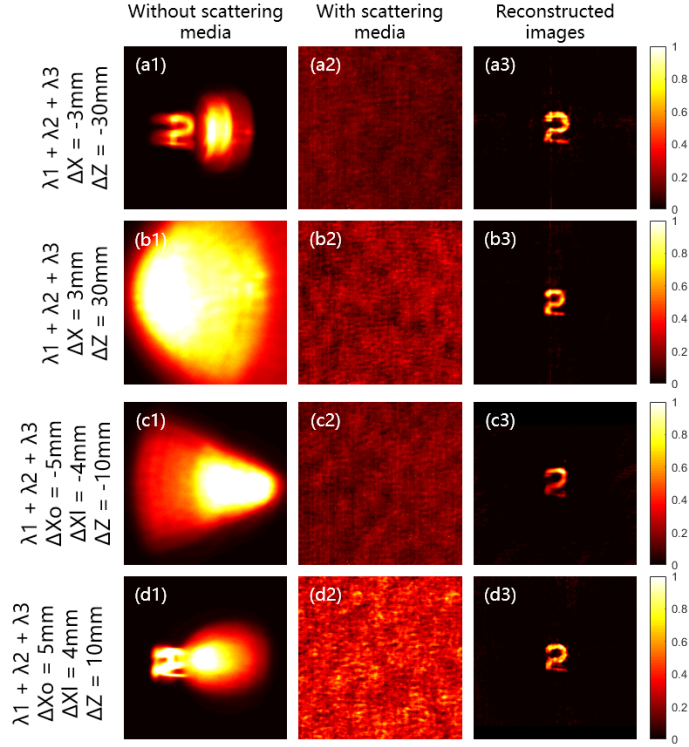

Figure. 6. Results with multiple aberrations at the condition of with or without scattering medium. (a1-d1) Direct images without scattering medium. (a2-d2) Speckle patterns with scattering medium. (a3-d3) Images reconstructed from speckle patterns by using phase-retrieval.

To achieve a global high-quality imaging in a traditional optical imaging system, complicated and expensive optical design is needed to eliminate kinds of optical aberrations. Compared with traditional imaging systems, we simply use a scattering medium to improve the imaging capabilities of the imaging systems on many parameters based on the optical memory effect. The proposed method can be performed non-invasive and in single-shot, while other imaging methods based on scattering media including optical phase conjugation [24, 25], speckle scanning [26], and point spread function (PSF) deconvolution [13, 27] cannot be achieved. However, due to the limitations of the ME and phase recovery algorithm, the improved imaging parameters are still limited by the ME range. As the development of imaging technology using scattering media, the quality of imaging will be better and better in the future. Scattering media can also achieve three-dimensional imaging [28], which further expands the role of scattering media in imaging systems.

## 5. Conclusion

In conclusion, we experimentally demonstrate the usage of scattering media in a simple single-lens imaging system to improve the imaging capabilities. By introducing a scattering medium in between the lens and camera, the original blurred images due to defocus, coma, spherical and chromatic dispersions can be recovered from the speckle patterns. Furthermore, all four important parameter ranges can be extended simultaneously. Compared with improving the imaging ability through expensive optical components and complex optical design, the proposed method by adding a scattering medium to the imaging systems is simple and cheap. Scattering media can be added to the original imaging system without changing any other optics, thus suited to many imaging systems. Although we only conduct experiments with four simple aberrations, it is also applicable to other types of aberrations because the random scattered light properties of the scattering medium can transmit the high-frequency spatial information lost by the imaging system to the imaging plane.

**Funding**

National Natural Science Foundation of China (NSFC) (61527821, 61905257, U1930115). Instrument Developing Project (YZ201538) and the Strategic Priority Research Program (XDB16) of the Chinese Academy of Sciences (CAS).

**Disclosures**

The authors declare no conflicts of interest.

**References**

1. K. E. Moore, "Algorithm for global optimization of optical systems based on genetic competition," in *Optical Design and Analysis Software*, (International Society for Optics and Photonics, 1999), pp. 40-47.
2. J. Rogers, "Using global synthesis to find tolerance-insensitive design forms," in *International Optical Design Conference 2006*, (SPIE, 2006).
3. M. Isshiki, D. Sinclair, and S. Kaneko, "Lens design: Global optimization of both performance and tolerance sensitivity.," in *International Optical Design Conference 2006*, (SPIE, 2006),
4. G. G. Gregory, M. Isshiki, J. M. Howard, D. C. Sinclair, S. Kaneko, and R. J. Koshel, "Lens design: global optimization of both performance and tolerance sensitivity," in *International Optical Design Conference 2006*, (2007).
5. M. van Turnhout, P. van Grol, F. Bociort, and H. P. Urbach, "Obtaining new local minima in lens design by constructing saddle points," Opt. Express **23**, 6679-6691 (2015).
6. R. Kingslake and R. Barry Johnson, *Lens Design Fundamentals (Second Edition)* (Academic Press, Boston, 2010).
7. D. Wu, H. Fan, J. Wang, and Y.-l. Li, "Multi-attribute automatic optimization method for lens system design," in *Fifth International Conference on Optical and Photonics Engineering*, (SPIE, 2017).
8. J. Bertolotti, E. G. Van Putten, C. Blum, A. Lagendijk, W. L. Vos, and A. P. Mosk, "Non-invasive imaging through opaque scattering layers," Nature **491**, 232-234 (2012).
9. O. Katz, P. Heidmann, M. Fink, and S. Gigan, "Non-invasive single-shot imaging through scattering layers and around corners via speckle correlations," Nat. Photonics **8**, 784-790 (2014).
10. Y. Choi, T. D. Yang, C. Fang-Yen, P. Kang, K. J. Lee, R. R. Dasari, M. S. Feld, and W. Choi, "Overcoming the diffraction limit using multiple light scattering in a highly disordered medium," Phys. Rev. Lett. **107**, 023902 (2011).
11. T. Nakamura, R. Horisaki, and J. Tanida, "Compact wide-field-of-view imager with a designed disordered medium," Opt. Rev. **22**, 19-24 (2015).
12. H. Yilmaz, E. G. van Putten, J. Bertolotti, A. Lagendijk, W. L. Vos, and A. P. Mosk, "Speckle correlation resolution enhancement of wide-field fluorescence imaging," Optica **2**(2015).
13. X. S. Xie, H. C. Zhuang, H. X. He, X. Q. Xu, H. W. Liang, Y. K. Liu, and J. Y. Zhou, "Extended depth-resolved imaging through a thin scattering medium with PSF manipulation," Sci Rep **8**, 4585 (2018).
14. M. Liao, D. Lu, G. Pedrini, W. Osten, G. Situ, W. He, and X. Peng, "Extending the depth-of-field of imaging systems with a scattering diffuser," Sci Rep **9**, 7165 (2019).
15. E. G. van Putten, D. Akbulut, J. Bertolotti, W. L. Vos, A. Lagendijk, and A. P. Mosk, "Scattering lens resolves sub-100 nm structures with visible light," Phys. Rev. Lett. **106**, 193905 (2011).
16. E. Edrei and G. Scarcelli, "Memory-effect based deconvolution microscopy for super-resolution imaging through scattering media," Sci Rep **6**, 33558 (2016).
17. P. Song, S. Jiang, H. Zhang, Z. Bian, C. Guo, K. Hoshino, and G. Zheng, "Super-resolution microscopy via ptychographic structured modulation of a diffuser," Opt. Lett. **44**, 3645-3648 (2019).
18. X. Shao, W. Dai, T. Wu, H. Li, and L. Wang, "Speckle-correlation imaging through highly scattering turbid media with LED illumination," in *Smart Biomedical and Physiological Sensor Technology XII*, (SPIE, 2015),
19. H. Li, T. Wu, J. Liu, C. Gong, and X. Shao, "Simulation and experimental verification for imaging of gray-scale objects through scattering layers," Appl. Optics **55**, 9731-9737 (2016).
20. P. Sheng, *Introduction to Wave Scattering, Localization, and Mesoscopic Phenomena* (Academic Press, 1995).
21. J. W. Goodman, *Speckle phenomena in optics: theory and applications* (Roberts and Company Publishers, 2007).
22. I. Freund, M. Rosenbluh, and S. Feng, "Memory effects in propagation of optical waves through disordered media," Phys. Rev. Lett. **61**, 2328-2331 (1988).
23. X. Xiao, S. Du, F. Zhao, J. Wang, J. Liu, and R. Li, "Single-shot optical speckle imaging based on pseudothermal illumination," Acta Phys. Sin. **68**, 034201 (2019).
24. Z. Yaqoob, D. Psaltis, M. S. Feld, and C. Yang, "Optical Phase Conjugation for Turbidity Suppression in Biological Samples," Nat. Photonics **2**, 110-115 (2008).
25. K. Wu, Q. Cheng, Y. Shi, H. Wang, and G. P. Wang, "Hiding scattering layers for noninvasive imaging of hidden

objects," Sci Rep **5**, 8375 (2015).
26. X. Yang, Y. Pu, and D. Psaltis, "Imaging blood cells through scattering biological tissue using speckle scanning microscopy," Opt. Express **22**, 3405-3413 (2014).
27. H. Zhuang, H. He, X. Xie, and J. Zhou, "High speed color imaging through scattering media with a large field of view," Sci Rep **6**, 32696 (2016).
28. R. Horisaki, Y. Okamoto, and J. Tanida, "Single-shot noninvasive three-dimensional imaging through scattering media," Opt. Lett. **44**, 4032-4035 (2019).